\documentclass[journal,twoside,web]{ieeecolor}
\usepackage{tmi}
\usepackage{cite}
\usepackage{amsmath,amssymb,amsfonts}
\usepackage{algorithmic}
\usepackage{graphicx}
\usepackage{textcomp}
\usepackage{dsfont}
\usepackage{color, soul}
\usepackage{diagbox}
\usepackage{makecell}

\def\BibTeX{{\rm B\kern-.05em{\sc i\kern-.025em b}\kern-.08em
    T\kern-.1667em\lower.7ex\hbox{E}\kern-.125emX}}
\usepackage{graphicx,amssymb,mathrsfs,amsmath,amsfonts,dsfont}
\usepackage{hyperref}
\usepackage{lineno}
\usepackage{stfloats}
\usepackage{algorithm}
\usepackage{algorithmic}
\usepackage{amsbsy}
\usepackage[mathscr]{euscript}
\usepackage{bm}
\usepackage{multirow}
\usepackage[utf8]{inputenc}

\newcommand{\x}{\bm{x}}

\newcommand{\y}{\bm{y}}
\newcommand{\z}{\bm{z}}

\newcommand{\n}{\bm{n}}

\newtheorem{thm}{Theorem}[section]

\newtheorem{rmk}{Remark}[section]
\newtheorem{assump}{Assumption}

\begin{document}
\title{ Meta-Learning Enabled Score-Based Generative Model for 1.5T-Like Image Reconstruction from 0.5T MRI}

\author{Zhuo-Xu~Cui, Congcong~Liu, Chentao~Cao, Yuanyuan~Liu, Jing~Cheng, Qingyong~Zhu, Yanjie~Zhu, Haifeng~Wang, Dong~Liang, \IEEEmembership{Senior Member, IEEE}
\thanks{This work was supported in part by the National Key R$\&$D Program of China (2020YFA0712202, 2017YFC0108802 and 2017YFC0112903); China Postdoctoral Science Foundation under Grant 2020M682990; National Natural Science Foundation of China (61771463, 81830056, U1805261, 81971611, 61871373, 81729003, 81901736); Natural Science Foundation of Guangdong Province (2018A0303130132); Shenzhen Key Laboratory of Ultrasound Imaging and Therapy (ZDSYS20180206180631473); Shenzhen Peacock Plan Team Program (KQTD20180413181834876); Innovation and Technology Commission of the government of Hong Kong SAR (MRP/001/18X); Strategic Priority Research Program of Chinese Academy of Sciences (XDB25000000).}
\thanks{Corresponding author:dong.liang@siat.ac.cn}
\thanks{Z.-X. Cui and C. Liu are contributed equally to this work.}
\thanks{Z.-X. Cui, Y. Liu, Q. Zhu and D. Liang are with Research Center for Medical AI, Shenzhen Institutes of Advanced Technology, Chinese Academy of Sciences, Shenzhen, China.}
\thanks{C. Cao, J. Cheng, H. Wang, Y. Zhu and D. Liang are with Paul C. Lauterbur Research Center for Biomedical Imaging, Shenzhen Institutes of Advanced Technology, Chinese Academy of Sciences, Shenzhen, China.}
}

\maketitle

\begin{abstract}
Magnetic resonance imaging (MRI) is known to have reduced signal-to-noise ratios (SNR) at lower field strengths, leading to signal degradation when producing a low-field MRI image from a high-field one. Therefore, reconstructing a high-field-like image from a low-field MRI is a complex problem due to the ill-posed nature of the task. Additionally, obtaining paired low-field and high-field MR images is often not practical. We theoretically uncovered that the combination of these challenges renders conventional deep learning methods that directly learn the mapping from a low-field MR image to a high-field MR image unsuitable. To overcome these challenges, we introduce a novel meta-learning approach that employs a teacher-student mechanism. Firstly, an optimal-transport-driven teacher learns the degradation process from high-field to low-field MR images and generates pseudo-paired high-field and low-field MRI images. Then, a score-based student solves the inverse problem of reconstructing a high-field-like MR image from a low-field MRI within the framework of iterative regularization, by learning the joint distribution of pseudo-paired images to act as a regularizer. Experimental results on real low-field MRI data demonstrate that our proposed method outperforms state-of-the-art unpaired learning methods.
\end{abstract}


\section{Introduction}
\label{sec:introduction}

\IEEEPARstart{M}{agnetic} resonance imaging (MRI) is a powerful tool in modern medicine, providing high-resolution images of the human body without invasive or radioactive procedures. However, the high magnetic field required by common hospital MRI scanners, which drives copper wire coils that must be actively cooled with liquid helium, makes these machines complex, expensive, and difficult to operate and maintain. This restricts the availability of high-field MRI technology to many doctors and patients worldwide, especially in developing countries. In contrast, low-field MRI prototypes typically employ permanent magnets, which are less expensive, simpler to operate and maintain, and thus have the potential to provide a more cost-effective and accessible alternative to high-field MRI scanners \cite{hori2021low,Gecmen2020}.
Although low-field MRI offers certain advantages such as reduced susceptibility artifacts and lower costs, its imaging resolution is limited by the low signal-to-noise ratio (SNR). This limitation can result in the incomplete visualization of detailed structures. To address this issue, improving the quality of low-field MRI is crucial for its wider acceptance. This paper aims to overcome this limitation by reconstructing high-field-like (1.5T-like) MR images from low-field (0.5T) MRI. 

\begin{figure}[!t]
\centering
\includegraphics[width=0.45\textwidth,height=0.25\textwidth]{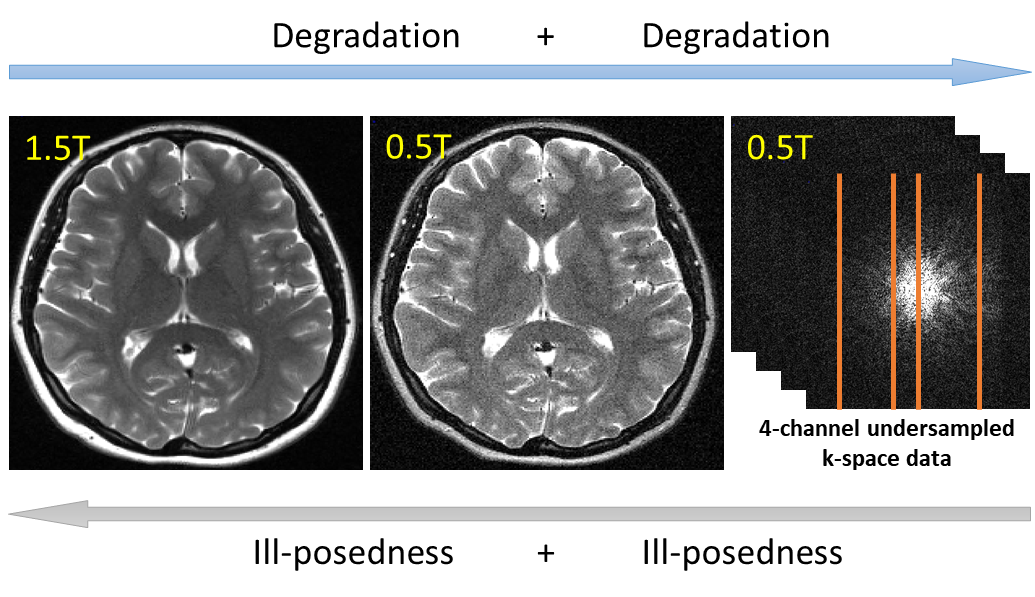}
\caption{The diagram depicts the dual degradation of 1.5T MR images to 0.5T undersampled k-space data, as well as the double ill-posed nature of its reversal.}
\label{f1}
\end{figure}

However, there are several challenges associated with this aim. Firstly, as the magnetic field strength decreases, both the SNR and resolution of MRI images exhibit a decreasing trend. Consequently, the transition from a high-field MRI to a low-field MRI entails a process of degradation, which means that reconstructing high-field-like MRI images from low-field MRI constitutes an ill-posed inverse problem. As we know, ill-posed problems are notoriously challenging due to their non-uniqueness and unstable nature \cite{Tikhonov1943OnTS,engl1996regularization,cui2022deep}. Therefore, accurately and stably obtaining the mapping from a low-field MRI image to a high-field MRI image presents a significant challenge, rendering traditional end-to-end deep learning methods inapplicable.

Secondly, most existing low-field MRI equipment features an open design, with the main magnetic field oriented perpendicular to the patient. The receiving coils are typically designed using a solenoid structure, and research indicates that using only four channels can achieve a high SNR while keeping costs reasonable \cite{marques2019low}. However, the low redundancy of the signal acquired by these four channels leads to strong ill-posedness of MRI reconstruction, adding to the difficulty of achieving the aim. These two challenges will lead to strong ill-posedness in reconstruction, as shown in Figure \ref{f1}.

Lastly, it is important to note that most existing high-field-like MR image reconstructions \cite{bahrami2016reconstruction,bahrami2016convolutional} often rely on learning from paired high-field and low-field MR images, which are challenging to obtain in practice. Alternatively, unpaired learning models have been suggested, but they focus solely on mapping low-field MR images to high-field MR images. This approach may not effectively address the inverse problem of reconstructing high-field-like MRI images from low-field MRI, as explained in the previous paragraph.

To overcome these challenges, it is imperative to devise a novel learning mechanism that can precisely capture the prior information of low-field and high-field MR images through unpaired data. Subsequently, a regularization model can be established based on this prior information to cope with the ill-posedness caused by reduced SNR and resolution, as well as low coil channel redundancy on low-field MRI.

\subsection{Contributions and Observations }
Based on these critical needs, this work makes several contributions and observations. 
\begin{enumerate}
\item To address the ill-posed challenge of reconstructing high-field-like MR images from low-field MRI in scenarios where paired high-field and low-field MR images are unavailable, we have devised a meta-learning approach utilizing a teacher-student learning mechanism. The proposed mechanism involves two stages. Firstly, an optimal-transport (OT) driven teacher is trained to approximate the degradation process from high-field to low-field MR images, thus generating pseudo-paired high-field and low-field MRI images. Subsequently, a score-based student is trained to solve the reconstruction inverse problem within the framework of iterative regularization. This is achieved by learning the joint distribution of the pseudo-paired high-field and low-field MR images, which serves as a regularizer.

\item Within the framework of OT theory, we prove that the mapping from low-field to high-field MR images in conventional unpaired learning does not exist under certain conditions. Conversely, we demonstrate that the degraded mapping from high-field to low-field MR images learned by our proposed teacher model is the unique solution of the OT model.

\item Thanks to the regularization framework described above, we are able to achieve fast reconstruction, i.e., reconstructing high-field-like images from undersampled low-field k-space data, which is difficult to achieve using existing methods.

\item Experiments on real low-field data demonstrate that the proposed method outperforms the state-of-the-art unpaired learning method in terms of reconstructed contrast, SNR and visualization of detailed structures.
\end{enumerate}

The remainder of the paper is organized as follows. Section \ref{sect_rw} reviews some related work. Section \ref{sect3} discusses the proposed teacher-student model based on meta-learning. The implementation details are presented in Section \ref{sect4}. Experiments performed on several data sets are presented in Section \ref{sect5}. The last section \ref{sect7} gives some concluding remarks. All the proofs are presented in the Appendix.

\section{Related Work}\label{sect_rw}
\subsection{Reconstruction of 7T MR Images from 3T MRI}
High-field MRI offers greater resolution and better tissue contrast in clinical settings, potentially enabling more accurate and early diagnosis of brain diseases \cite{wattjes2009high,van2013clinical}. Nevertheless, its high cost and technical complexity restrict its accessibility to many doctors and patients worldwide, especially in developing countries. As a result, generating high-quality images similar to those obtained with high-field MRI from low-field MRI, which is more affordable and easier to use, has important practical implications.

In the initial stages of research, the primary objective was to approach the challenge of reconstructing high-field MRI images from low-field MRI as a super-resolution problem. Essentially, researchers aimed to reconstruct high-resolution images (from high-field MRI) using low-resolution images (from low-field MRI) \cite{bahrami20177t,de2022deep}. However, the degradation that occurs when transitioning from high-field MRI to low-field MRI is not limited solely to resolution. Issues such as reduced signal-to-noise ratio and changes in contrast must also be addressed.
To tackle these challenges, \cite{bahrami2016reconstruction} developed a paired dictionary based on paired 3T and 7T MR images. This dictionary was used to represent the relationship between 3T and 7T MR images and ultimately achieve the reconstruction of MR images that are comparable in quality to those produced by 7T MRI scanners, using 3T MRI.

Compared to reconstructing 7T MR images from 3T MR, the task tackled in this paper poses greater difficulties due to the following factors: Firstly, the imaging SNR reduces by around 2.3 times from 7T to 3T, whereas the drop is roughly threefold from 1.5T to 0.5T \cite{bernstein2006field}, leading to a more severe degradation of the image quality and ill-posedness of the reconstruction problem. Secondly, the signal acquisition coil's redundancy is lower for 0.5T MRI than for 3T MRI, exacerbating the reconstruction problem's ill-posedness. Thirdly, Reference \cite{bahrami2016reconstruction}'s approach requires paired 3T and 7T MR training images, whereas this paper aims to address a more practical scenario where paired training images are unavailable. Hence, the direct application of Reference \cite{bahrami2016reconstruction}'s method is challenging, and an unpaired learning mechanism is proposed to extract and refine the prior information of 1.5T and 0.5T MR images as regularization to deal with the ill-posedness.
\subsection{Unpaired Learning}
In reality, obtaining paired medical images is difficult due to incidental factors such as motion and scanner switches. Therefore, extracting correlation priors from unpaired data has become a popular research direction. There are two approaches to address this problem. The first involves generating paired data based on physical principles, such as using Bloch simulation to generate low-field MR images from high-field MR images, and then utilizing paired learning methods. However, since physical principles typically involve a series of simplified approximations, the generated images often become distorted \cite{https://doi.org/10.1002/mrm.29077}. The second approach involves directly establishing a model that does not require paired training data, such as OT-driven WGAN \cite{NEURIPS2018_91d0dbfd}, cycleGAN \cite{Zhu_2017_ICCV,sim2020optimal,9173689}, etc. If cycleGAN or similar models are applied to the reconstruction task considered in this paper, it will learn the mapping between 0.5T and 1.5T MR images from unpaired data. However, it's important to note that the mapping from 1.5T to 0.5T MR images represents a degradation process, and we will demonstrate that the transmission mapping between 0.5T and 1.5T MR images does not exist, making it challenging to directly apply such methods.
\subsection{Meta-Learning}
Meta-learning is a robust learning framework that involves the concept of "learning to learn" \cite{zou2022meta}. Its primary application lies in the realm of few-shot learning. Traditional deep learning methods rely on mapping measurement data to labels, which can be problematic when label data is limited. In such cases, meta-learning offers a way for models to learn new knowledge from just a few examples, enabling them to learn by themselves \cite{pmlr-v70-finn17a}.

In \cite{9059750}, meta-learning was introduced to the field of low-dose CT imaging, where only a small amount of labeled data is available. The authors adopted a teacher-student learning paradigm, where the teacher, acting as a meta-learner, is trained on the small amount of labeled data and generates pseudo-labels for the unlabeled low-dose CT data. The student, acting as a base-learner, then maps the low-dose CT data to the pseudo-labels generated by the teacher.

However, due to the ill-posed nature resulting from low-field MRI, the mapping learning approach becomes insufficient, and a student with stronger representation ability based on data prior is necessary to address this issue. Additionally, the semi-supervised learning approach utilized by the teacher is not viable for the unpaired data scenario presented in this paper. Based on this inspiration, a novel meta-learning model that is suitable for the scenario under consideration will be proposed.
\subsection{Score-Based Diffusion Models}
The diffusion model is a powerful tool that allows for the integration and extraction of data priors, and has demonstrated successful applications across numerous fields \cite{yang2022diffusion,kazerouni2022diffusion,croitoru2023diffusion}. In this approach, the original data is subjected to a gradual perturbation process involving Gaussian noise, achieved through the use of stochastic partial differential equations (SDEs) \cite{song2021scorebased,chung2022score,cao2022high,cui2023spirit} or Markov chains \cite{NEURIPS2020_4c5bcfec} . 
This perturbation continues until the data is fully encoded as random noise, with a score-matching method \cite{vincent2011connection} employed to learn the prior distribution on the trajectory of this perturbation. Finally, the noise is decoded back into the original data by means of executing reverse SDEs or reverse Markov chains.

The diffusion model can extract a prior distribution that proves to be a valuable regularization term in MRI reconstruction applications. Specifically, the undersampled k-space data can be used to perform conditional inverse SDE or Markov chain, which serves as an iterative algorithm to solve the imaging regularization model. This paper delves into the capacity of diffusion model integration to extract prior knowledge and characterizes the correlation prior between 1.5T and 0.5T MRI images using a joint distribution. By utilizing this joint distribution as a regularization term, a reconstruction regularization model is constructed, enabling the successful reconstruction of 1.5T-like MRI images from undersampled 0.5T MRI data.

\begin{table}[!t]
  \caption{\label{Symbol Meaning}Mathematical symbols and their meaning.}\label{math:s}
  \centering
  \footnotesize
      \begin{tabular}{c|c}
        \hline \hline Symbol & Symbol Meaning \\
        \hline  
        $\x^{1.5}$ or $\x^{0.5}$ & 1.5T or 0.5T MR image \\
        $\y^{1.5}$ or $\y^{0.5}$ & 1.5T or 0.5T undersampled k-space data \\
        $\x^H$ & Hermitian transpose of $\x$, i.e., conjugate and transpose\\
        $p(\x)$ & probability density function (P.D.F.) of $\x$ \\
        $p(\x,\y)$ & joint P.D.F. of $\x$ and $\y$\\
        $\delta_{\x}$ & Dirac measure on $\x$ \\
        $\alpha$ or $\beta$ & distributions that $\x^{1.5}$ or $\x^{0.5}$ obeys, respectively\\
        $\bm{T}$ & measurable map\\
        $\bm{T}_\sharp\alpha$ & the image measure of $\alpha$ through the map $\bm{T}$\\
        \hline \hline
    \end{tabular}
\end{table}
\section{Methodology}\label{sect3}
In this section, we begin by introducing a forward model that maps 1.5T MR images to undersampled k-space data at 0.5T. We then propose a meta-learning approach using teacher-student mechanism, where the teacher learns the forward model from unpaired data to generate pseudo-paired data, and the student learns the joint distribution of the paired data to construct a regularized model for reconstructing 1.5T MR images from under-sampled k-space data at 0.5T.
\subsection{Forward Model of 1.5T Image Degradation to 0.5T MR Data}
The degradation of 1.5T MR images to 0.5T MR images can be described by a forward model expressed as:
\begin{equation}\label{eq:1}
\x^{0.5} = \bm{f}(\x^{1.5})
\end{equation}
Here, $\x^{0.5}$ and $\x^{1.5}$ are the 0.5T (low-field) and 1.5T (high-field) MR images, respectively, and $\bm{f}$ is an unknown nonlinear degradation process that maps high-field images to low-field images. On the other hand, the acquisition of 0.5T k-space data can be modeled as:
\begin{equation}\label{eq:2}
\y^{0.5} = \bm{A}\x^{0.5} + \bm{n}
\end{equation}
Here, $\y^{0.5}$ is the undersampled 0.5T k-space data, $\bm{A}$ is the encoding matrix, and $\bm{n}$ is Gaussian measurement noise. In the case of multichannel acquisition, $\bm{A=PFS}$, where $\bm{S}$ denotes the coil sensitivities, $\bm{F}$ represents the Fourier transform, and $\bm{P}$ denotes the undersampling pattern.

Typically, 0.5T MRI systems use only 4 coils to measure k-space data, making (\ref{eq:2}) a strongly ill-posed problem. Moreover, to reconstruct a 1.5T MR image, one needs to solve the inverse problem (\ref{eq:1}). Thus, the ill-posedness of reconstructing a 1.5T-like image from 0.5T undersampled k-space data is more severe than the general fast MRI problem.

\begin{figure*}[thbp]
\begin{center}
\includegraphics[width=0.85\textwidth,height=0.25\textwidth]{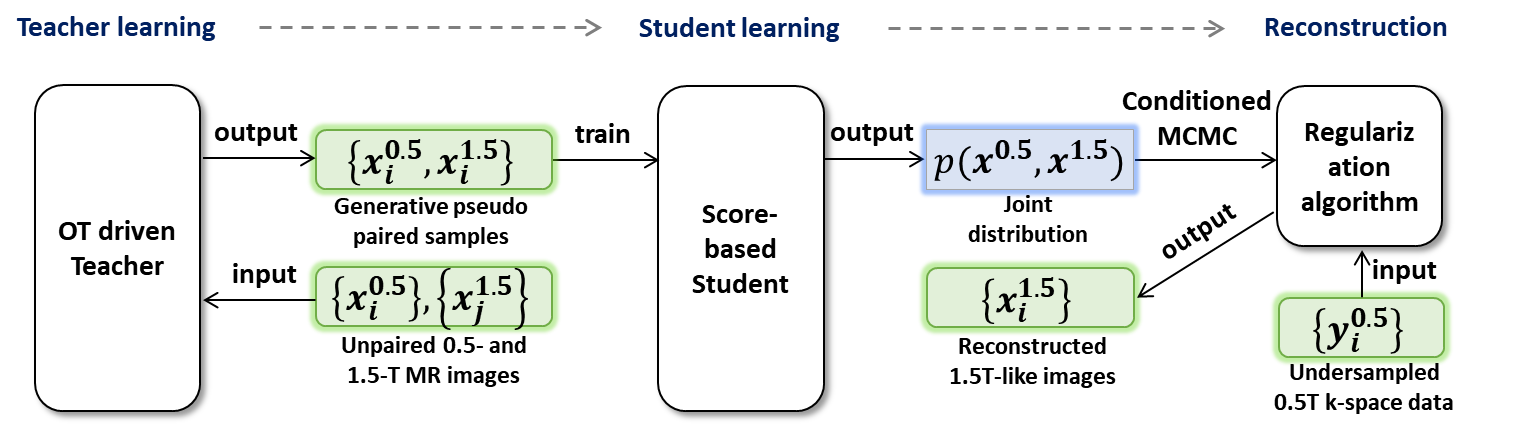}
\end{center}
\caption{Illustration of 1.5T-like image reconstruction from 0.5T MRI using meta-learning framework. (a) Firstly, an OT driven teacher learns the degradation process from high-field to low-field MR images using unpaired data $\{\x_i^{0.5}\}, \{\x_j^{1.5}\}$, generating pseudo-paired high-field and low-field MRI images $\{\x_i^{0.5}, \x_i^{1.5}\}$. (b) Then, by learning the joint distribution $p(\x_i^{0.5}, \x_i^{1.5})$ to act as a regularizer, a score-based student solves the inverse problem of reconstructing a 1.5T-like MR image from undersampled 0.5T k-space data $\{\y_i^{0.5}\}$ within the framework of iterative regularization.}
\label{f1}
\end{figure*}
\subsection{OT Driven Teacher}
To tackle the inverse problems presented in equations (\ref{eq:1}) and (\ref{eq:2}), traditional end-to-end deep learning approaches often rely on approximating the mapping from $\x^{0.5}$ to $\x^{1.5}$ using neural networks. However, in scenarios where paired data is not available, such as the one explored in this paper, GAN-based methods can be used to overcome the need for paired data by elevating the point-to-point loss function to the loss function between data distributions. Specifically, the GAN-based method seeks an operator $\bm{T}^{\beta\rightarrow\alpha}$ that satisfies the condition in equation \begin{equation} \label{b2a}
    \bm{T}^{\beta\rightarrow\alpha}_\sharp(\beta)=\alpha
\end{equation} 
where $\x^{1.5}\sim\alpha$ and $\x^{0.5}\sim\beta$, and $\alpha,\beta$ denote data distribution. By doing so, the operator $\bm{T}$ is able to successfully achieve $\bm{T}^{\beta\rightarrow\alpha}(\x^{0.5})=\x^{1.5}$.

In practice, we encounter difficulty in finding an accurate $\bm{T}^{\beta\rightarrow\alpha}$ that can effectively map $\x^{0.5}$ to $\x^{1.5}$. Although we will provide numerical validation in the experimental section, we will also demonstrate theoretically that such a $\bm{T}^{\beta\rightarrow\alpha}$ does not exist. As a result, we must explore alternative methods for solving inverse problems (\ref{eq:1}) and (\ref{eq:2}). In this paper, we propose a roundabout learning approach that involves first learning the mapping $\bm{f}$ and then applying regularization to address the inverse problems (\ref{eq:1}) and (\ref{eq:2}).

Our proposed approach assumes the existence of a training set $\{\x_{j}^{1.5}\}_{j=1}^N$ and $\{\x_{j}^{0.5}\}_{j=1}^N$, where $\{\x_{j}^{0.5}\}_{j=1}^N$ is generated by shuffling the elements of $\{\bm{f}(\x_{j}^{1.5})\}_{j=1}^N$. Before presenting our method, we make the following critical assumptions:
\begin{assump}\label{ass:1}
Each element in the training set $\{\x_{j}^{1.5}\}_{j=1}^N$ is unique, meaning that
$\x_i^{1.5}\neq\x_j^{1.5}$
for $i\neq j$.
Since (\ref{eq:1}) is ill-posed, its solution is not unique. Thus, it is assumed that there exists at least one $\x_j^{0.5}$ with solutions $\x_i^{1.5}$ and $\x_{i'}^{1.5}$ ($i\neq i'$), i.e.
$$\x_j^{0.5}=\bm{f}(\x_i^{1.5})=\bm{f}(\x_{i'}^{1.5}).$$
\end{assump}
\begin{assump}\label{ass:2}
There exists a cost function $c$ such that, for any $i\in[N]$,  
$$c(\bm{f}(\x^{1.5}_i),\x^{1.5}_i)\leq c(\bm{f}(\x^{1.5}_j),\x^{1.5}_i),~ j\neq i.$$
\end{assump}
\begin{rmk}
In simpler terms, Assumption \ref{ass:2} suggests that MR images of the same individual or slice, but captured with different magnetic field strengths, exhibit greater similarity than those taken at different field strengths from different individuals or slices.    
\end{rmk}
\begin{thm}\label{thm:1}
We introduce discrete 1.5T and 0.5T MR image distributions, denoted by $\alpha$ and $\beta$ respectively, defined as follows: 
$$\alpha=\frac{1}{N}\sum_{i=1}^N\delta_{\x^{1.5}_i},~\beta=\frac{1}{N}\sum_{i=1}^N\delta_{\x^{0.5}_i}$$
Assuming that Assumptions \ref{ass:1} and \ref{ass:2} hold, we can draw the following conclusions:
\begin{enumerate}
\item It is not possible to find a mapping (\ref{b2a}) that can transfer the distribution of low-field MR images $\beta$ to that of high-field MR images $\alpha$.
\item The OT problem of finding a mapping from the 1.5T MR image distribution $\alpha$ to the 0.5T MR image distribution $\beta$,
\begin{equation}\label{ot}
    \bm{T}^{\alpha\rightarrow\beta}=\arg\min_{\bm{T}}\left\{\sum_{i=1}^Nc(\bm{T}(\x_{i}^{1.5}),\x_{i}^{1.5}):\bm{T}_{\sharp}\alpha=\beta\right\}
\end{equation}
has a unique solution $\bm{T}^{\alpha\rightarrow\beta}=\bm{f}$.
\end{enumerate}
\end{thm}
\begin{rmk}
Our first conclusion provides theoretical support for our previous claim that using an end-to-end deep learning method to directly map 0.5T MR images to 1.5T MR images (\ref{b2a}) is no longer viable due to the problem of ill-posedness. On the other hand, our second conclusion shows that even when using unpaired training data, it is possible to learn the degradation process from 1.5T MR images to 0.5T MR images accurately by employing the OT model (\ref{ot}).
\end{rmk}

It is noteworthy that (\ref{ot}) is often difficult to compute, and if $c(\cdot,\cdot)=\|\cdot-\cdot\|$, (\ref{ot}) can be relaxed to the following Kantorovich dual form \cite{MAL-073}.
\begin{equation}\label{dual}
\bm{T}^{\alpha\rightarrow\beta}=\arg\min_{\bm{T}}\max_{\varphi\in \text{Lip}_1}\int\varphi(\bm{T}(\z))d\beta(\z)- \int\varphi(\x)d\alpha(\x)
\end{equation}
where $\text{Lip}_1$ denotes the collection of functions that are uniformly Lipschitz continuous with constant 1, which can often be achieved through spectral normalization \cite{miyato2018spectral}. Numerically, we can obtain $\bm{T}^{\alpha\rightarrow\beta}$ and maximizer $\varphi^*$ through an alternating maximization-minimization iterative optimization algorithm. Once we obtain the optimal $\bm{T}^{\alpha\rightarrow\beta}$, we can execute the following algorithm to generate pseudo-samples paired with 1.5T and 0.5T.

\begin{algorithm}[htb]
	\caption{$\text{Pseudo-Paired Data Generator (Teacher)}$.}
	\label{alg:2}
	\begin{algorithmic}[1]
		\STATE {\bfseries Input:} $\bm{T}^{\alpha\rightarrow\beta}$ and 1.5T MR images $\{\x_i^{1.5}\}^N_{i=1}$;\\

	\FOR{$i=1,2,\ldots,N$}
	\STATE $\x^{0.5}_i= \bm{T}^{\alpha\rightarrow\beta}(\x^{1.5}_i)$
        \ENDFOR
	\STATE {\bfseries Output:} $\{(\x_i^{0.5},\x_i^{1.5})\}^N_{i=1}$.
	\end{algorithmic}
\end{algorithm}

\subsection{Score-Based Student}
Due to the accuracy of the optimization process (\ref{dual}) in obtaining $\bm{f}$, it is conventional to combine the nonlinear inverse problems (\ref{eq:1}) and (\ref{eq:2}) into the following equation:
$$ \y^{0.5} = \bm{A}\bm{f}(\x^{1.5}) + \bm{n}$$
From a statistical point of view, solving the inverse problem above can be achieved through posterior probability sampling, which involves calculating the gradient of the following function:
\begin{equation}\label{score-mri}
 \log p(\widetilde{\x}^{1.5}|\y^{0.5})=\log p(\y^{0.5}|\widetilde{\x}^{1.5}) + \log p(\widetilde{\x}^{1.5}) + C
\end{equation}
In the case of regularization, the prior distribution $\log p(\widetilde{\x}^{1.5})$ acts as a regularizer. For diffusion models, we perturb clean 1.5T image ${\x}^{1.5}$ using a sequence of Gaussian noise with different scales and approximate $\nabla\log p(\widetilde{\x}^{1.5})$ on the perturbation trajectory using the score-matching method, where $\widetilde{\x}^{1.5}$ denotes the perturbed image. However, due to the nonlinearity of $\bm{f}$ (i.e., $\bm{T}^{\alpha\rightarrow\beta}$), it is challenging to determine what distribution type $\bm{f}$ will map the perturbation noise in $\widetilde{\x}^{1.5}$ to, making it difficult to compute $\log p(\y^{0.5}|\widetilde{\x}^{1.5})$. This issue motivated us to propose a new method for reconstructing ${\x}^{1.5}$ from $\y^{0.5}$.

To avoid the computation of $\log p(\y^{0.5}|\widetilde{\x}^{1.5}_t)$, we propose the following posterior probability model to solve inverse problems (\ref{eq:1}) and (\ref{eq:2}) without the need to involve their combination form:
\begin{equation}\label{like}\log p(\widetilde{\x}^{1.5},\widetilde{\x}^{0.5}|\y^{0.5})=\log p(\y^{0.5}|\widetilde{\x}^{0.5}) + \log p(\widetilde{\x}^{0.5},\widetilde{\x}^{1.5}) + C
\end{equation}
As per the forward model (\ref{eq:2}), the data consistency yields
$\nabla \log p(\y^{0.5}|\widetilde{\x}^{0.5})=-\frac{\bm{A^H}(\bm{A}\x_{t}-\y)}{\gamma^2+\varepsilon^2}$,
where $\gamma$ represents the noise scale of $\n$ in (\ref{eq:2}) and $\varepsilon$ represents the scale of
perturbed noise in $\widetilde{\x}$. With the pseudo-paired data generated by Algorithm \ref{alg:1}, we can approximate the joint distribution $p(\widetilde{\x}^{1.5},\widetilde{\x}^{0.5})$ using the score-matching method.

More specifically, we perturb $X_0:=[\x^{0.5},\x^{1.5}]$ by Gaussian noise with scales $\{\varepsilon_i\}_{i=1}^T$ that satisfy $\varepsilon_1<\varepsilon_2<\cdots<\varepsilon_T$. Let $p(\widetilde{X}_i,|X_0)=\mathcal{N}(\widetilde{X}_i|X_0,\varepsilon_i^2\bm{I})$, and the perturbed data distribution is $p(\widetilde{X}_i)=\int p(\widetilde{X}_i|X_0)\text{d}\pi(X_0)$, where $\pi:=(\bm{T}^{\alpha\rightarrow\beta},\bm{I})_{\sharp}\alpha$. If $\varepsilon_1$ is chosen such that $\varepsilon_1$ approximates 0, then $p_{\varepsilon_1}(X)=p(X_0)$ holds.

Based on the score-matching method, we can estimate $\nabla \log p(\widetilde{X}_i)$ at all the scales by training a joint score function $\bm{s}_{\bm{\phi}}(\widetilde{X}_i,\varepsilon_i)$ with the following loss function:
\begin{small}\begin{equation}\label{score_match}\bm{\phi}^*=\arg\min_{\bm{\phi}}\frac{1}{2L}\sum_{i=1}^{L}\mathbb{E}_{\pi(X_0)}\mathbb{E}_{p(\widetilde{X}_i,|X_0)}\left[\left\|\varepsilon_i\bm{s}_{\bm{\phi}}(\widetilde{X}_i,\varepsilon_i)+\frac{\z}{\varepsilon_i}\right\|^2\right]\end{equation}\end{small}
where $\z_t\sim \mathcal{N}(\bm{0},\bm{I})$.

Once we have trained a scoring function, we can carry out conditionally Langevin MCMC sampling based on equation (\ref{like}), that is
\begin{equation*}\begin{aligned}
\widetilde{X}_{i+1}&=\widetilde{X}_{i}+\frac{\eta_i}{2}\nabla \log p(\widetilde{X}_{i}|\y^{0.5})+\sqrt{\eta_i}\z_i\\
&=\widetilde{X}_{i}+\frac{\eta_i}{2}(\nabla \log p(\widetilde{X}_{i})+\nabla \log p(\y^{0.5}|\widetilde{\x}^{0.5}_{i}))+\sqrt{\eta_i}\z_i\\
&=\widetilde{X}_{i}+\frac{\eta_i}{2}\left(\bm{s}_{\bm{\phi}}(\widetilde{X}_{i},\varepsilon_{i})-\frac{\bm{A^H}(\bm{A}\widetilde{\x}_{i}^{0.5}-\y^{0.5})}{\gamma^2+\varepsilon_i^2}\right)+\sqrt{\eta_i}\z_i
\end{aligned}\end{equation*}
In particular, the choice of parameter $\eta_i$ follows that of literature \cite{NEURIPS2020_92c3b916}, and the conditional Langevin MCMC sampling is detailed in Algorithm \ref{alg:2}.
\begin{algorithm}[htb]
	\caption{Conditional Langevin MCMC Sampling (Student).}
	\label{alg:1}
	\begin{algorithmic}[1]
		\STATE {\bfseries Input:} $\y^{0.5}$, $\bm{\phi}^*$, $\{\varepsilon_i\}_{i=1}^L$, $\epsilon$ and $T$;\\
		\STATE {\bfseries Initialize:} $X_0:=[\widetilde{\x}^{0.5}_0,\widetilde{\x}^{1.5}_0]$;\\
		\FOR{$i=L,L-1,\ldots,1$}
        \STATE $\eta_i=\epsilon\cdot\varepsilon_i^2/\varepsilon^2_L$
        \FOR{$t=0,2,\ldots,T-1$}
		\STATE Draw $\z_t\sim \mathcal{N}(\bm{0},\bm{I})$:
\begin{footnotesize}\begin{equation*}
\widetilde{X}_{t+1}=\widetilde{X}_{t}+\frac{\eta_i}{2}\left(\bm{s}_{\bm{\phi}^*}(\widetilde{X}_{t},\varepsilon_{i})-\frac{\bm{A^H}(\bm{A}\widetilde{\x}^{0.5}_{t}-\y^{0.5})}{\gamma^2+\varepsilon_i^2}\right)+\sqrt{\eta_i}\z_t
\end{equation*}\end{footnotesize}
		\ENDFOR
        \STATE $\widetilde{X}_0=\widetilde{X}_T$
        \ENDFOR
		\STATE {\bfseries Output:} $\widetilde{X}_T=[\widetilde{\x}^{0.5}_T,\widetilde{\x}^{1.5}_T]$.
	\end{algorithmic}
\end{algorithm}

\section{Implementation}\label{sect4}
The evaluation was performed on MRI data from SuperMark 1.5T scanner and SuperMark 0.5T scanner. 
\subsection{Data Acquisition}
\subsubsection{Anke 0.5T MRI data}
The brain raw data was collected using a 0.5T Anke scanner (SuperMark 0.5T) with a 4-channel receive coil and conventional Cartesian 2D FSE protocol for both T1-weighted and T2-weighted scans. The imaging parameters were as follows: TR/TE = 450/14 ms (T1W), TR/TE = 3000/120 ms (T2W), FOV of $230\times 230$ mm, a slice thickness of 5 mm, single average, matrix size of $320\times 320$, resolution of $0.7 \times 0.7$, and bandwidth of 130 Hz/Pixel. 

\subsubsection{Anke 1.5T MRI data}
The brain raw data was collected using a 1.5T Anke scanner (SuperMark 1.5T) with a 8-channel receive coil and conventional Cartesian 2D FSE protocol for both T1-weighted and T2-weighted scans. The imaging parameters were as follows: TR/TE=410/10 ms (T1W), TR/TE=3000/100 ms (T2W), FOV of $230\times 230$ mm, a slice thickness of 5 mm, single average, matrix size of $320\times 320$, resolution of $0.7 \times 0.7$, and bandwidth of 240 Hz/Pixel.

We obtained data from 20 individuals using 1.5T and 0.5T scanners, capturing 19 slices of data from each individual. Of these, data from 18 individuals was utilized as the training set. It is important to highlight that the images collected using the 1.5T scanner were not paired with those collected using the 0.5T scanner. To ensure ease of evaluating the quality of reconstruction, we carefully selected data from two individuals whose 1.5T and 0.5T images were relatively paired, to form the test set.

\subsection{Network Architecture and Training}
To train the teacher model, we used the ResNet with 9 blocks to represent the transport operator $\bm{T}$, following the methodology proposed in \cite{Zhu_2017_ICCV}. For $\varphi$ in the OT loss function (\ref{dual}), we employed the pyramid discriminator network. We optimized the loss function (\ref{dual}) using the ADAM optimizer with $\beta_1=0.5, \beta_2=0.999$, following the same approach as in \cite{Zhu_2017_ICCV}. The mini-batch size was set to 1, and we trained the model for 200 epochs using a learning rate of $10^{-3}$ with an ExponentialLR scheduler with $\gamma=0.95$.

To train the student model, we utilized the NCSNv2 network \cite{NEURIPS2020_92c3b916} to learn the score function. The specific parameters used were $\varepsilon_L=50$, $\varepsilon_1=0.0066$, and the number of classes was set to 266. We enabled the Exponential Moving Average (EMA) with an EMA rate of 0.999. We used the ADAM optimizer with $\beta_1=0.9, \beta_2=0.999$ to optimize the loss function (\ref{score_match}). The mini-batch size was set to 1, and we trained the model for 500 epochs using a learning rate of $10^{-4}$.

We implemented our models on an Ubuntu 20.04 operating system equipped with an NVIDIA A100 Tensor Core (GPU, 80 GB memory) using the open PyTorch 1.10 framework \cite{paszke2019pytorch} with CUDA 11.3 and CUDNN support.

\subsection{Performance Evaluation}
In this study, our primary method of evaluating the quality of reconstructed images from 0.5T data to 1.5T-like MR images relies on visual analysis, as there is no paired test set available for these two image types. To provide a more objective evaluation, we conduct registration of the 1.5T MR images with the 0.5T MR images, which allows us to calculate quantitative metrics such as peak signal-to-noise ratio (PSNR) and normalized mean square error (NMSE). These metrics enable us to quantify the differences in SNR and contrast between the reconstructed images and the true 1.5T images. However, it is important to note that these quantitative metrics should be interpreted with caution, as they can be affected by the blurring caused by registration and inherent structural differences between the 1.5T and 0.5T images that are not paired. Therefore, while these metrics serve as a useful reference, they cannot provide a definitive evaluation.

\section{Experimentation Results}\label{sect5}
In this section, our objective is to assess the efficacy of our proposed method, Meta-Score-MRI, and to validate our above claims through a series of experiments. Firstly, we will compare our approach with the mapping learning method CycleGAN \cite{Zhu_2017_ICCV} to support our assertion that the latter is unsuitable for this particular scenario. Moreover, we will approximate $p(\y^{0.5}|\widetilde{\x}^{1.5})$ in model (\ref{score-mri}) by using a Gaussian distribution, which necessitates the approximation of $\bm{f}$ as a noisy process. This approximation, known as Score-MRI, will enable us to demonstrate the superiority of our proposed model (\ref{like}).

During the experiments, we will initially ablate the issue of ill-posedness caused by undersampling (\ref{eq:2}). This will allow us to evaluate the performance of generating fully sampled 0.5T MR images to 1.5T MR images. Subsequently, we will examine the impact of dual ill-posedness by considering 3x undersampling on 0.5T data. Since 0.5T data has only 4 channels, 3-fold undersampling is close to the limit. This will help us validate the performance of various methods under these conditions.

\subsection{Reconstruction of 1.5T-like MR Image from Full-Sampled 0.5T MRI Data}
\begin{figure*}[!t]
\centering
\includegraphics[width=0.96\textwidth,height=0.5\textwidth]{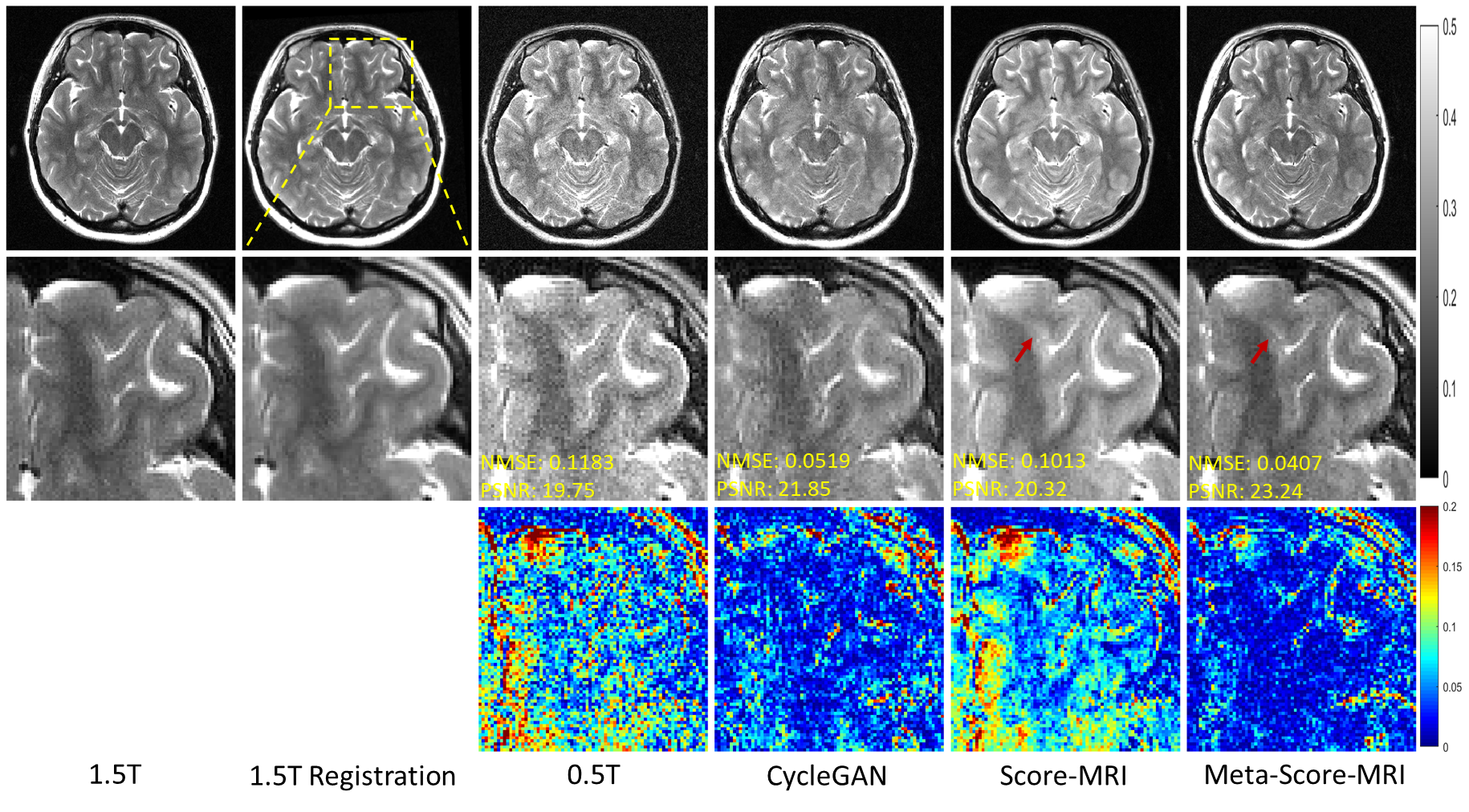}
\caption{Reconstruction of 1.5T-like T2W MR Images from full-sampled 0.5T T2W MRI.
The first line showcases the original 1.5T MR image, while the second line demonstrates the registration of the 1.5T image with the shape of the 0.5T MR image. The third line showcases the 0.5T MR image. The resulting reconstructed images are depicted in lines four through six. On the other hand, the second row offers an enlarged view, and the third row shows the error view between the generated 1.5T-like image and the registered 1.5T image. The NMSE/PSNR values of the enlarged view are displayed in the corners. The grayscale of the reconstructed images is visible on the right side of the figure.}
\label{f2}
\end{figure*}

This section presents the testing of various methods on full-sampled 0.5T MRI data, with the aim of reconstructing 1.5T-like MR images and eliminating the ill-posedness outlined in (\ref{eq:2}). The results of the different methods used in reconstructing 1.5T-like T2W images are shown in Figure \ref{f2}. The images obtained from the 0.5T MRI are observed to have reduced SNR and contrast degradation when compared to the 1.5T images.

Among the reconstruction methods employed, it was discovered that CycleGAN yielded results with similar contrast to actual 1.5T images. However, magnified images revealed visible artifacts, thus validating Theorem \ref{thm:1}, which postulated that there is no OT mapping from 0.5T to 1.5T images. On the other hand, the Score-MRI method only considered SNR degradation in $\bm{f}$, as captured in model (\ref{score-mri}), while ignoring contrast degradation. As a result, its experimental outcomes were in line with its modeling, with improved image SNR, but no change in contrast relative to the 0.5T images.

The proposed method, Meta-Score-MRI, reconstructed 1.5T-like images that were visually almost indistinguishable from the real 1.5T images in terms of SNR and contrast. Notably, Score-MRI's modeling was inaccurate, leading to blurry image details, as indicated by the red arrow. In contrast, the proposed method accurately reconstructed these details, further validating the superiority of the proposed method's modeling captured in equation (\ref{like}).

\begin{figure*}[!t]
\centering
\includegraphics[width=0.96\textwidth,height=0.5\textwidth]{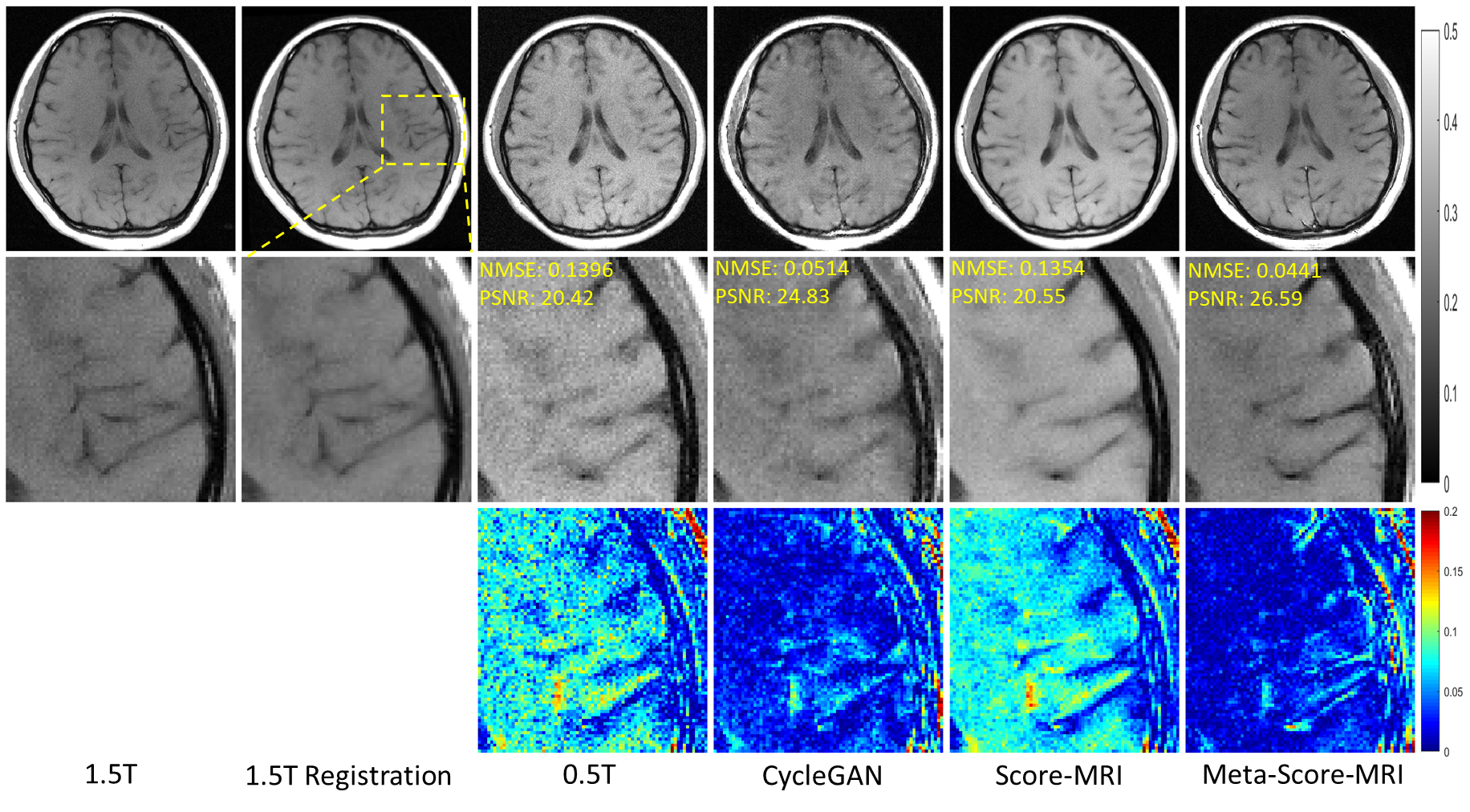}
\caption{Reconstruction of 1.5T-like T1W MR Images from full-sampled 0.5T T1W MRI.
The first line showcases the original 1.5T MR image, while the second line demonstrates the registration of the 1.5T image with the shape of the 0.5T MR image. The third line showcases the 0.5T MR image. The resulting reconstructed images are depicted in lines four through six. On the other hand, the second row offers an enlarged view, and the third row shows the error view between the generated 1.5T-like image and the registered 1.5T image. The NMSE/PSNR values of the enlarged view are displayed in the corners. The grayscale of the reconstructed images is visible on the right side of the figure.}
\label{f3}
\end{figure*}

The results of different methods for reconstructing 1.5T T1W images are presented in Figure \ref{f3}. It is evident that the image reconstructed by CycleGAN displays grid-like artifacts. While Score-MRI performs consistently with the previous experiment by enhancing the image's SNR, the contrast level remains at that of a 0.5T level. Additionally, the sulcus region of the Score-MRI outcome appeared blurred. In contrast, the proposed method shows a significant improvement in both contrast and SNR, resulting in a clear reconstruction in the sulcus region that matches the quality of 1.5T T1W images.

To ensure an impartial evaluation of the reconstruction quality, we have included Table \ref{tab:1a}, which showcases the quantitative metrics that were obtained by comparing the reconstructed results with registered 1.5T images. While it's worth noting that the structure of the registered 1.5T images may not be entirely consistent with that of the reconstructed images, these metrics still provide valuable insights into the reconstruction's SNR and contrast. The quantitative metrics reveal that the proposed method outperforms other comparable techniques, offering further confirmation of its effectiveness.

\begin{table}
	\begin{center}
		\caption{Quantitative comparison for various methods on reconstructing 1.5T-like MR image from full-sampled 0.5T MRI.}\label{tab:1a}
		\setlength{\tabcolsep}{2.0mm}{
			\begin{tabular}{l|l|cc}
				\hline
				\multicolumn{ 2}{c}{ Datasets} & \multicolumn{ 2}{|c}{Quantitative Evaluation}  \\
				\multicolumn{ 2}{c|}{ \& Methods   } &NMSE &PSNR(dB)  \\
				\hline
				\multirow{4}{*}{T2W Recon}
				& 0.5T &0.1772$\pm$0.0667&20.29$\pm$1.70 \\
				\cline{2-4}
				& CycleGAN  &0.1537$\pm$0.0456&21.11$\pm$1.53 \\
				\cline{2-4}
				& Score-MRI  &0.1564$\pm$0.0533&21.11$\pm$1.92 \\
				\cline{2-4}
				& Meta-Score-MRI  &\textcolor{red}{0.1476$\pm$0.0380}&\textcolor{red}{21.58$\pm$1.41}\\
				\hline
				\multirow{4}{*}{T1W Recon}
				& 0.5T  &0.2023$\pm$0.0730&20.72$\pm$1.79 \\
				\cline{2-4}
				& CycleGAN  &0.0986$\pm$0.0423&24.44$\pm$1.99 \\
				\cline{2-4}
				& Score-MRI  &0.0987$\pm$0.0343&23.75$\pm$1.86 \\
				\cline{2-4}
				& Meta-Score-MRI  &{0.0987$\pm$0.0392}&\textcolor{red}{25.07$\pm$1.94}\\
				\hline
		\end{tabular}}
	\end{center}
\end{table}

\subsection{Reconstruction of 1.5T-like MR Image from Undersampled 0.5T MRI Data}
\begin{figure*}[!t]
\centering
\includegraphics[width=0.96\textwidth,height=0.5\textwidth]{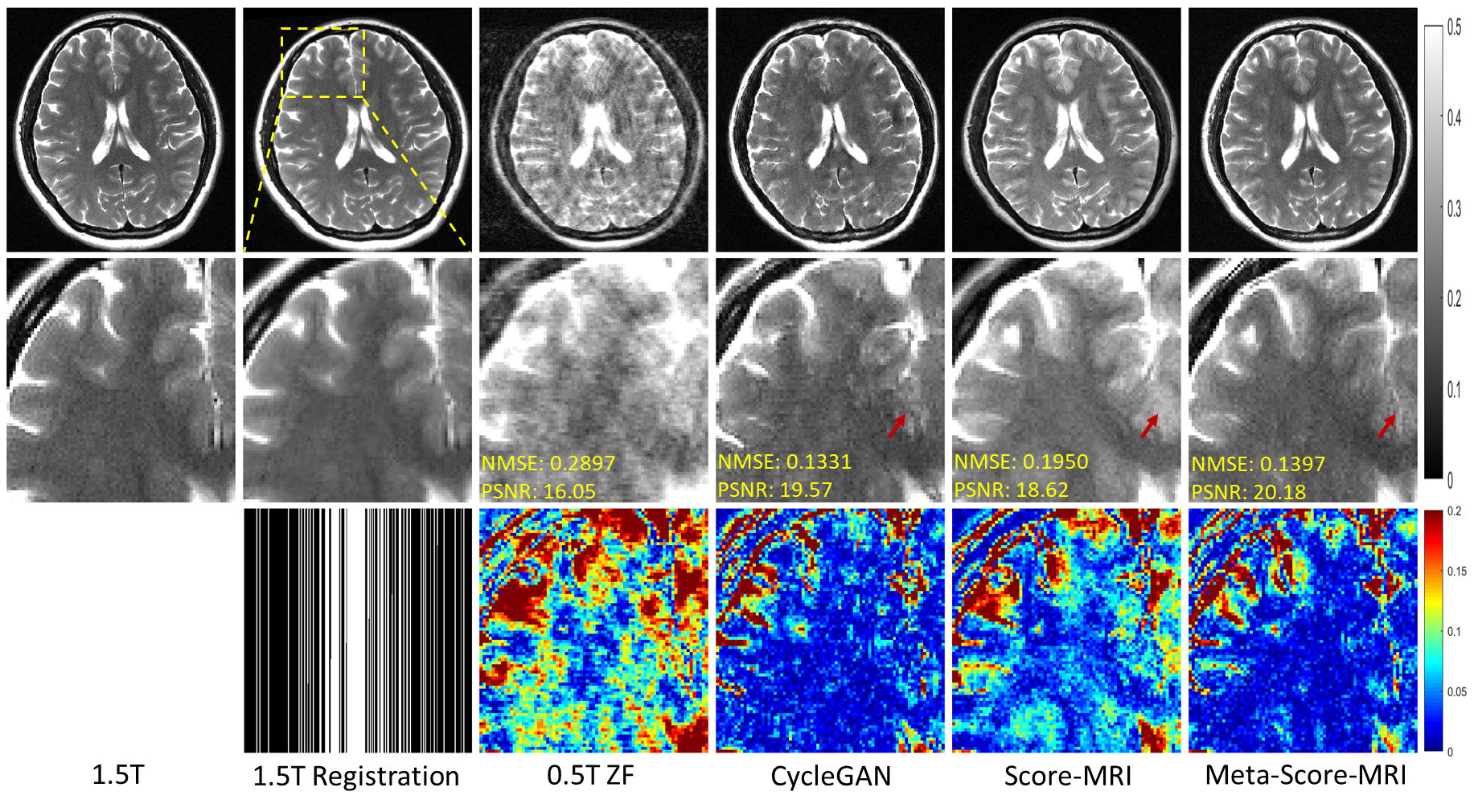}
\caption{Reconstruction of 1.5T-like T2W MR Images from undersampled 0.5T T2W k-space data under acceleration rate $R=3$.
The first line showcases the original 1.5T MR image, while the second line demonstrates the registration of the 1.5T image with the shape of the 0.5T MR image. The third line showcases the zero-filled 0.5T MR image. The resulting generated images are depicted in lines four through six. On the other hand, the second row offers an enlarged view, and the third row shows the error view between the generated 1.5T-like image and the registered 1.5T image. The NMSE/PSNR values of the enlarged view are displayed in the corners. The grayscale of the reconstructed images is visible on the right side of the figure.}
\label{f4}
\end{figure*}

In this section, our focus is on reconstructing a 1.5T-like MR image from 3-fold undersampled 0.5T k-space data, and evaluating the performance of various methods for solving the dual ill-posed problems presented in equations (\ref{eq:1}) and (\ref{eq:2}). As shown in Figure \ref{f4}, we present the reconstruction results of different methods from undersampled 0.5T T2W data. It is worth noting that the 0.5T zero-filled image exhibits a high noise level and noticeable aliasing patterns, making this experiment more challenging than the previous one.

Specifically, images reconstructed by CycleGAN exhibit aliasing patterns that obscure image details, while both Score-MRI and the proposed method, which leverages a score-based prior, effectively suppress the aliasing patterns. However, the contrast in the images produced by Score-MRI remains at the level of the 0.5T image. In contrast, the proposed method generates results that closely resemble the true 1.5T image in terms of SNR and contrast. Furthermore, the proposed method accurately reconstructs image details, including the brain sulcus indicated by the red arrow, which was not properly reconstructed by Score-MRI due to its modeling inaccuracies.

\begin{figure*}[!t]
\centering
\includegraphics[width=0.96\textwidth,height=0.5\textwidth]{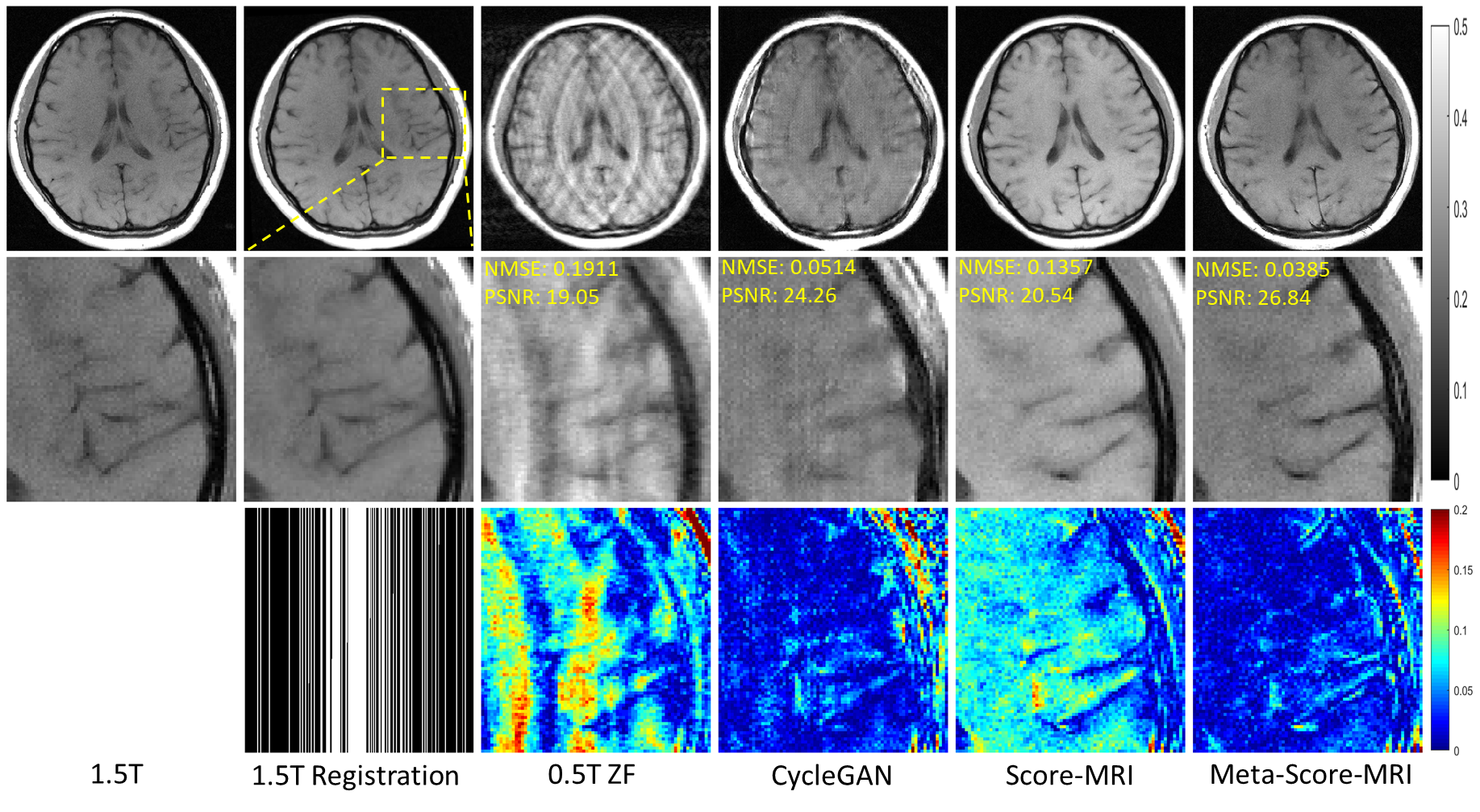}
\caption{Reconstruction of 1.5T-like T1W MR Images from undersampled 0.5T T1W k-space data under acceleration rate $R=3$.
The first line showcases the original 1.5T MR image, while the second line demonstrates the registration of the 1.5T image with the shape of the 0.5T MR image. The third line showcases the zero-filled 0.5T MR image. The resulting generated images are depicted in lines four through six. On the other hand, the second row offers an enlarged view, and the third row shows the error view between the generated 1.5T-like image and the registered 1.5T image. The NMSE/PSNR values of the enlarged view are displayed in the corners. The grayscale of the reconstructed images is visible on the right side of the figure.}
\label{f5}
\end{figure*}

The results of the reconstruction performed on 3-fold undersampled 0.5T T1W k-space data are presented in Figure \ref{f5}. The outcomes obtained with CycleGAN reveal clear signs of aliasing patterns, which obscure image details. Conversely, Score-MRI successfully mitigates these patterns; however, the contrast of the reconstructed images remains at the level of 0.5T images, resulting in blurry image details. In contrast, the proposed method not only suppresses aliasing artifacts but also achieves the same level of SNR, contrast, and imaging details as 1.5T T1W images in the reconstructed results.

Table \ref{tab:1b} displays the quantitative metrics comparing the reconstruction results of all methods with the registered 1.5T images. It is abundantly clear that the proposed method outperforms the comparison methods by a significant margin in terms of the quantitative metrics. This observation highlights the effectiveness of the proposed method in addressing the dual ill-posedness stemming from low field strength and accelerated imaging.
\begin{table}
	\begin{center}
		\caption{Quantitative comparison for various methods on reconstructing 1.5T-like MR image from 3x undersampled 0.5T MRI.}\label{tab:1b}
		\setlength{\tabcolsep}{2.0mm}{
			\begin{tabular}{l|l|cc}
				\hline
				\multicolumn{ 2}{c}{ Datasets} & \multicolumn{ 2}{|c}{Quantitative Evaluation}  \\
				\multicolumn{ 2}{c|}{ \& Methods   } &NMSE &PSNR(dB)  \\
				\hline
				\multirow{4}{*}{T2W Recon}
				& 0.5T ZF  &0.2147$\pm$0.0723&19.43$\pm$1.71 \\
				\cline{2-4}
				& CycleGAN  &0.1652$\pm$0.0392&20.72$\pm$1.31 \\
				\cline{2-4}
				& Score-MRI  &0.1578$\pm$0.0534&21.11$\pm$1.86 \\
				\cline{2-4}
				& Meta-Score-MRI  &\textcolor{red}{0.1467$\pm$0.0365}&\textcolor{red}{21.68$\pm$1.51}\\
				\hline
				\multirow{4}{*}{T1W Recon}
				& 0.5T ZF &0.2363$\pm$0.0965&20.18$\pm$2.16 \\
				\cline{2-4}
				& CycleGAN  &0.1048$\pm$0.0324&23.60$\pm$1.60 \\
				\cline{2-4}
				& Score-MRI  &0.0964$\pm$0.0355&23.89$\pm$1.93 \\
				\cline{2-4}
				& Meta-Score-MRI  &\textcolor{red}{0.0949$\pm$0.0384}&\textcolor{red}{25.00$\pm$1.98}\\
				\hline
		\end{tabular}}
	\end{center}
\end{table}

\section{Conclusion}\label{sect7}
This paper presented a model for reconstructing high-field-like MR images from low-field MRI, which poses a dual-ill-posed problem with unpaired data. Theoretical analysis and experimental verification demonstrated that conventional mapping-learning methods are no longer suitable for this task. To address this issue, we proposed a teacher-student model based on meta-learning, which solves the dual-inverse problem from a regularization perspective. The experimental results indicated that our proposed method is capable of reconstructing MR images from undersampled 0.5T k-space data, with image quality metrics such as SNR, contrast, and detail imaging comparable to those of 1.5T MRI images.

Low-field MRI is an inexpensive and easily accessible device that offers greater convenience to patients and doctors in underdeveloped areas. However, its widespread use is limited by the degradation of imaging SNR and contrast. The proposed method allows for the reconstruction of MRI images with low-field MRI that are comparable to those produced by high-field MRI. Consequently, the adoption of this method is expected to have a considerable social impact.
\appendix
\subsection{Proof of Theorem \ref{thm:1}}\label{app:1}
If there exists a $\bm{T}_{\sharp}\beta=\alpha$, then the equation holds 
\begin{equation}\label{ot:eq}
\frac{1}{N}=\sum_{j:T(\x^{0.5}_j)=\x^{1.5}_s}\frac{1}{N}
\end{equation}
for any $s\in[N]$. On the other hand, Assumption \ref{ass:1} implies that $\x_j^{0.5}=\bm{f}(\x_i^{1.5})=\bm{f}(\x_{i'}^{1.5})=\x_{j'}^{0.5}$. If $\bm{T}$ maps $\x_j^{0.5}$ to $\x_s^{1.5}$, then there must be another index $j'$ such that $\bm{T}(\x^{0.5}_{j'})=\x^{1.5}_s$. This leads to a contradiction since it implies $\frac{1}{N}=\frac{2}{N}$.

Conversely, it is straightforward to verify that $\bm{f}_{\sharp}\alpha=\beta$, meaning that $\bm{f}$ satisfies \eqref{ot:eq}. On the other hand, assuming there exists an optimal $\bm{T}$ such that $\sum_{i=1}^Nc(\bm{T}(\x_{i}^{1.5}),\x_{i}^{1.5})\leq \sum_{i=1}^Nc(\bm{f}(\x_{i}^{1.5}),\x_{i}^{1.5})$ contradicts Assumption \ref{ass:2}. Therefore, we have $\bm{T = f}$.

\section*{Acknowledgement}
The authors would like to express their gratitude to Shenzhen Anke High-tech Co., Ltd. for providing the 0.5T and 1.5T MRI data necessary for this paper, which were instrumental in validating the proposed methods.

\bibliographystyle{ieeetr}
\bibliography{library_manu}

\end{document}